\documentstyle[aps,12pt,tighten,amssymb,epsf,cite]{revtex}

\title{\vspace*{2.0in} Logarithmic Operators
Fold $D$ branes into AdS$_3$ }
\vspace{1.0in}
\author{ {\bf John Ellis}$^{a}$, 
{\bf N.E.~Mavromatos}$^{a,b}$ 
and {\bf Elizabeth Winstanley}$^{b}$ }
\vspace{1.0in}
\address{${}^a$Theory Division, CERN, CH--1211
  Geneva 23, Switzerland, \\
${}^b$Theoretical Physics (University of Oxford), 1 Keble
  Road, Oxford, OX1 3NP, U.K.}

\begin{document}

\maketitle

\vspace*{2cm}
\begin{centering} 

{\bf Abstract} 

\end{centering}
\begin{abstract}
We use logarithmic conformal field theory techniques 
to describe recoil effects in the scattering of two Dirichlet branes
in $D$ dimensions. In the
particular case that a $D1$ brane strikes a $D3$ brane
perpendicularly, thereby folding it, we find that the recoil space-time is
maximally symmetric, with AdS$_3 \otimes {\rm E}_{D-3}$ geometry. 
We comment on
the possible applications of this result to the study of
transitions between different background metrics.
\end{abstract}

\vspace*{-6in}
\begin{flushright}
CERN--TH/99--279\\
OUTP--99--42P \\
hep--th/9909068
\end{flushright}
\vspace*{6in}

\newpage
Non-perturbative techniques have provided many new insights
into the theory formerly known as strings~\cite{SR}. In particular,
a rich variety of classical soliton solutions have been
found using the technology of Dirichlet ($D$) branes~\cite{Polch}. These
must be taken
into account in any study of possible ground states and the
spectra of excitations. A crucial ingredient in such a study
is understanding the interactions between $D$ branes and
`elementary' closed-string states, and with other $D$ branes.
In general, since recoil perturbs the classical $D$-brane solutions, 
one might expect that such interactions, would
involve departures from conformal symmetry and criticality.

One example of this recoil phenomenon was given in~\cite{kanti98},
where the perturbation of a $D$ particle by 
interaction with a closed-string state in flat space was considered.
In the low-energy limit, where the closed-string energy $E \ll M_D$, the
mass of the $D$ particle, it was shown that the 
background metric was perturbed to an anti-de-Sitter (AdS)
form~\cite{ellis98}.
Moreover, the perturbation was described by a
logarithmic pair of world-sheet operators corresponding to the position
and velocity of the struck $D$ brane, and it was shown that momentum
was conserved during the interaction and recoil process.

The purpose of this paper is to extend and generalize this 
previous discussion by
considering a generic interaction between two extended $D$ branes that do
not have particle interpretations. 
Such constructions may be relevant in the context of  
viewing the observable world as a $D3$ brane
embedded in a higher-dimensional 
space-time, with the extra `bulk' dimensions being transverse to the 
$D3$-brane coordinates~\cite{dimopo98,antoniadis98,antoniadis99}.  

\begin{figure}[htb]
\begin{center}
\epsfxsize=4in
\bigskip
\centerline{\epsffile{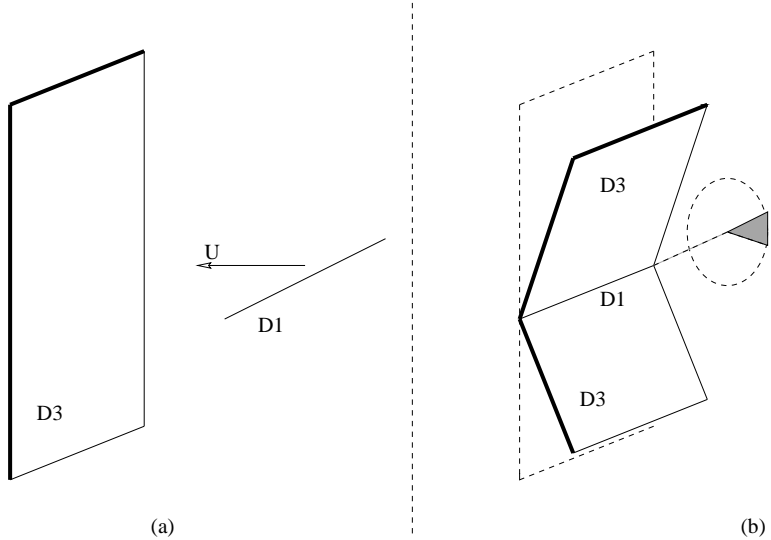}}
\vspace{0.2in}
\caption{\it Schematic representation of the folding effect
in $D$-brane/$D$-brane collisions:
(a) a $D1$ brane 
moving with velocity $U$ along a `bulk' direction 
perpendicular to a 
$D3$ brane embedded in a 
$D$-dimensional Euclidean
space time E$_{D}$ strikes the $D3$ brane (b), which is then folded,
and the space-time around it is distorted 
into AdS$_3 \otimes {E}_{D-3}$. The dashed circle 
around the $D1$ direction
in (b) indicates the angular deficit that appears when 
the bulk direction along which the $D1$ brane
was moving is compactified to a circle. 
\label{fig1}}
\end{center}
\end{figure}

Remarkably, we find just one case
in which the recoil space has an AdS geometry, namely AdS$_3$, which
arises when a macroscopic string ($D1$ brane) moving along 
some bulk direction in space-time hits a $Dp$ brane perpendicular
to the direction of motion, as shown in Fig.~\ref{fig1}.
This physical problem is analogous to a straight stick hitting a
flat sheet of paper, which we would expect to fold after the impact.
As in the closed-string/$D$-particle scattering case
considered previously, the
$D$-brane/$D$-brane interaction is described by a logarithmic pair
of world-sheet operators. In this case, they 
do indeed describe folding of the
struck $D$ brane. We are also able to demonstrate that momentum
is conserved during the interaction, as in the previous
closed-string/$D$-particle scattering case.

There have been many previous studies of string theory on an AdS$_3$
background~\cite{electrovac,ads3lit}, 
which can be described using a critical conformal
field theory. Our analysis opens the way to studies of transitions
between this and other string backgrounds. It confirms that such
transitions are described by non-conformal logarithmic field theories.
We also recall that the boundary of the AdS$_3$ background is equivalent
to an appropriate coset of the $SL(2,R)$ manifold that
appears in the analysis of two-dimensional black holes in string
theory~\cite{WittenBH}. Perhaps the mathematical analysis 
presented here could be used as the basis for
studies of the formation of such black-hole backgrounds, much as
transitions between different two-dimensional string black holes have
already been described in terms of logarithmic pairs of world-sheet
operators~\cite{ellis96}. However, the study of such a possibility lies
beyond
the scope of this article.
\vskip0.3cm

We first review the world-sheet formalism 
based on logarithmic operators that was developed
in a series of papers~\cite{kogan96,ellis96,mavro+szabo}, 
for the mathematical description of the recoil of a 
$D$ brane when struck by a closed-string state or by another $D$ brane.
Although closed-string/$D$-brane scattering has 
been studied using this formalism, we are unaware of any
treatment of $D$-brane/$D$-brane scattering in
the relevant literature so far. 
Logarithmic conformal field theory~\cite{lcft} lies on the border
between finite conformal field theories
and general renormalizable two--dimensional quantum field theories.
It is the relevant tool~\cite{kogan96,ellis96,mavro+szabo} 
for this problem, because
the recoil process involves a change 
of state (transition) in the string background, and as such 
is not described by a conformal field theory.
This change of state induced by the recoil process 
can be described a change in the
$\sigma$--model background, and as such is a non--equilibrium process.
This is reflected~\cite{ellis96,mavro+szabo} in the logarithmic
operator algebra itself.

As discussed in references~\cite{kogan96,ellis96,mavro+szabo} in the
case of $D$--brane string solitons, their recoil after interaction with
a closed-string (graviton) state is characterized 
by a $\sigma$ model deformed by a pair of logarithmic
operators~\cite{lcft}:
\begin{equation}
C^I_\epsilon = \epsilon \Theta_\epsilon (X^I),\qquad 
D^I_\epsilon = X^I \Theta_\epsilon (X^I), \qquad I \in \{0,\dots, 3\}
\label{logpair}
\end{equation} 
defined on the boundary $\partial \Sigma$ of the string
world sheet. Here $X^I, I\in \{0, \dots, 3\}$ obey Neumann boundary
conditions on the
string world sheet, and denote the brane coordinates.  The remaining
$y^i, i\in \{4, \dots, 9\}$ denote the transverse 
bulk directions.  In the case of $D$ particles, which were 
examined in~\cite{kogan96,ellis96,mavro+szabo}, 
the index $I$ takes the value $0$
only, in which case the operators (\ref{logpair}) act as deformations
of the conformal field theory on the world sheet. The operator $U_i
\int _{\partial \Sigma} \partial_n X^i D_\epsilon $ describes the
movement of the $D$ brane induced by the scattering, where $U_i$
is its
recoil velocity, and $Y_i \int _{\partial \Sigma} \partial_n X^i
C_\epsilon $ describes quantum fluctuations in the initial position
$Y_i$ of the $D$ particle. 
It has been shown rigorously~\cite{mavro+szabo}  
that the logarithmic conformal algebra ensures 
energy--momentum  conservation during the recoil process: 
$U_i = \ell_s 
g_s ( k^1_i +
k^2_i)$, where $k^1 (k^2)$ is the momentum of the propagating closed
string state before (after) the recoil,  
and $g_s$ is the string coupling,
which is assumed here to be weak enough to ensure that $D$ branes are
very massive, with mass $M_D=1/(\ell _s g_s)$, where $\ell _s$ is the
string length. 

The correct specification of the logarithmic pair in equation
(\ref{logpair2}) entails a regulating parameter
$\epsilon\rightarrow0^+$, which appears inside the $\Theta_\epsilon
(t)$ operator:
\[\Theta_\epsilon (X^I) = \int \frac{d\omega}{2\pi}\frac{1}{\omega
-i\epsilon} e^{i\omega X^I} .\] In order to realize the logarithmic
algebra between the operators $C$ and $D$, one takes~\cite{kogan96}:
\begin{equation} 
\epsilon^{-2} \sim \ln [L/a] \equiv \Lambda,
\label{defeps}
\end{equation} 
where $L$ ($a$) are infrared (ultraviolet) world--sheet cutoffs.  The
recoil operators (\ref{logpair2}) are slightly relevant, in the sense
of the renormalization group for the world--sheet field theory, having
small conformal dimensions $\Delta _\epsilon = -\epsilon^2/2$.
Thus the $\sigma$ model perturbed by these operators is not conformal
for $\epsilon \ne 0$, and the theory requires Liouville
dressing~\cite{david88,distler89,ellis96}. Momentum
conservation is assured when the Liouville field is identified with
the time variable.
\vskip0.3cm

In the case of $Dp$ branes, the pertinent deformations are slightly
more complicated. As discussed in~\cite{kogan96}, the
deformations are given by 
\begin{equation} 
\sum_{I} g^D_{Ii} \int _{\partial \Sigma}
\partial_n X^i D_\epsilon ^I \qquad{\mathrm and}\qquad\sum_{I}
g^C_{Ii} \int _{\partial \Sigma} \partial_n X^i C^I_\epsilon~.
\label{logpair2}
\end{equation} 
The
$0i$ components of the two-index couplings
$g^{\alpha}_{Ii},~\alpha\in\{C,D\}$ include the collective momenta and
coordinates of the $D$ brane as in the $D$--particle case above, but now
there are additional couplings $g^{\alpha}_{Ii},~I\ne0$, which describe
the folding of the $D$ brane.
Such a folding may be caused by scattering with another macroscopic
object, namely another $D$ brane,  
propagating in a transverse direction, as shown schematically
in Fig.~\ref{fig1} for the 
case of a $D1$ brane hitting a $D3$ brane.
This situation is the most interesting
to us, since it generates an AdS$_3$ space,
as we show below. For symmetry reasons,
in the situation depicted in 
Fig. \ref{fig1}, the folding of the $D3$ brane 
occurs symmetrically around the axis of the $D1$ brane.
In this case, the precise logarithmic operator deformations 
shown in (\ref{logpair2}), which pertain only 
to the spatial region $y_i > 0$ for the Dirichlet coordinates, 
should be supplemented with their counterparts for 
the $y_i < 0$ region as well. This would, in principle, require
additional $\Theta (\pm y_i)$ factors, which would complicate the 
analysis without introducing any new points of principle. 
Therefore, for simplicity, we
restrict ourselves here to the $y_i > 0$  patch of space-time,
away from the hypersurface $y_i=0$. This will be implicit
in what follows.  

The folding couplings $g_{Ii}^D \equiv g_{Ii},~I\in\{0, \dots, p\},
~i\in\{p+1, \dots, 9\}$, are
relevant couplings with world--sheet renormalization--group
$\beta$ functions of the form
\begin{equation} 
  \beta_{g_{Ii}} = \frac{d}{d t} g_{Ii} = -\frac{1}{2t} g_{Ii} , \qquad 
t \sim \epsilon ^{-2}~. 
\label{betaf}
\end{equation} 
This implies that one may construct an exactly marginal set of
couplings ${\overline g_{Ii}}$ by redefining
\begin{equation}
{\overline g_{Ii}} \equiv \frac{g_{Ii}}{\epsilon}~.
\label{marginal}
\end{equation}
The renormalized couplings ${\overline g_{0i}}$ 
were shown in~\cite{mavro+szabo} to play the
r\^ole of the physical recoil velocity of the $D$ brane, while the
remaining
${\overline g_{Ii}},~I\ne 0$, describe the folding of the $Dp$ brane
for $p\ne 0$.

As discussed in~\cite{ellis96,kanti98}, 
the deformations (\ref{logpair}) 
create a local distortion of
the space-time surrounding the recoiling folded $D$ brane,
which may be determined using the
method of Liouville dressing.   
In~\cite{ellis96,kanti98} we concentrated on 
describing the resulting space-time 
in the case when a $D$ particle, embedded in a $D$-dimensional
space time, recoils after the scattering of a closed string
off the $D$--particle defect. 
To leading order in the recoil velocity $u_i$ 
of the $D$ particle, the resulting space-time
was found, for times $t \gg 0$
long after the scattering event at
$t=0$, to be equivalent to a Rindler wedge,
with apparent `acceleration' $\epsilon u_i$~\cite{kanti98}, where
$\epsilon$ is defined above (\ref{defeps}). 
For times $t < 0$, the space-time is flat Minkowski~\footnote{There is
hence a discontinuity at $t =0$, which leads to particle
production and decoherence for a low-energy spectator field theory 
observer who performs local scattering experiments
long after the scattering, and far away from the 
location of the collision of the closed string with the 
$D$ particle~\cite{kanti98}.}.
\vskip0.3cm

We now discuss the generalization of this phenomenon
to the case of single $D$-brane/$D$-brane scattering. 
For definiteness, we first consider the case where there are 
$y_i,~i=1, \dots m$ transverse (bulk) dimensions (with Dirichlet 
boundary conditions), and $X^I,~I=m+1, \dots D-1$ 
are longitudinal coordinates on the brane,
which obey Neumann boundary conditions.
In this generic case, we consider the folding 
of the recoiling brane along several directions, 
determined by couplings $g_{Ii}$ of appropriate logarithmic operators
of the form (\ref{logpair2}), and the recoil-velocity vector of the
$D$ brane along the $i$'th bulk direction is denoted by $u_i$.  
As mentioned previously, we restrict ourselves
for simplicity to the region of space-time in which $y_i >0$. 

The folding/recoil deformations of the $Dp$ brane (\ref{logpair2})
are relevant deformations, with anomalous 
dimension  $-\epsilon^2/2 $, which disturbs the conformal 
invariance of the $\sigma$ model, and restoration of conformal 
invariance requires Liouville dressing~\cite{distler89}. 
To determine the effect of such dressing on the space-time 
geometry, it is essential  to write~\cite{ellis96} the 
boundary recoil deformations 
as a bulk world-sheet deformations
\begin{equation}
    \int _{\partial \Sigma} {\overline g}_{Iz} x\Theta_\epsilon (x) 
\partial_n z =
\int _\Sigma \partial_\alpha \left({\overline g}_{Iz}^1 x\Theta_\epsilon (x) 
\partial ^\alpha z \right) 
\label{a1}
\end{equation}
where the ${\overline g}_{Iz}$ denote the renormalized
folding/recoil couplings (\ref{marginal}), 
in the sense discussed in~\cite{mavro+szabo}.
As we have already mentioned, such couplings are marginal on a flat
world sheet. From now on, for notational simplicity, we
rename ${\overline g}_{Ii} \rightarrow g_{Ii}$. 
The operators (\ref{a1}) are marginal also on a curved  
world sheet, provided~\cite{distler89} one 
Liouville-dresses the (bulk) integrand by multiplying it
by a factor $e^{\alpha_{Ii}}\phi$, where $\phi$ is the Liouville field
and $\alpha_{Ii}$ is the gravitational conformal dimension, which 
is related to the flat-world-sheet anomalous dimension $-\epsilon^2/2$ of
the recoil operator, 
viewed as a bulk world-sheet deformation, as follows~\cite{distler89}:
\begin{equation}
\alpha_{Ii}=-\frac{Q_b}{2} + 
\sqrt{\frac {Q_b^2}{4} + \frac {\epsilon^2}{2} } 
\label{anom}
\end{equation}
where $Q_b$ is the central-charge deficit of the bulk world-sheet theory.
In the recoil prblem at hand, as discussed in~\cite{kanti98},
$Q_b^2 \sim \epsilon^4/g_s^2$ for weak folding deformations $g_{Ii}^1$. 
This yields $\alpha_{Ii} \sim -\epsilon $ to leading order in 
perturbation theory in $\epsilon$, to which we restrict ourselves here. 

We next remark that, as the analysis of~\cite{ellis96} indicates, 
the $X^I$-dependent field operators 
$\Theta_\epsilon (X^I)$ scale as follows with $\epsilon$:  
$\Theta_\epsilon(X^I) \sim e^{-\epsilon X^I}
\Theta(X^I)$, where $\Theta(X^I)$ is a Heavyside step function
without any field content, evaluated in the limit $\epsilon \rightarrow 0^+$. 
The bulk deformations, therefore, yield the following 
$\sigma$-model terms:
\begin{equation} 
\frac{1}{4\pi \ell_s^2} \epsilon \sum_{I=m+1}^{D-1} g_{Ii} X^I 
e^{\epsilon(\phi_{(0)} - X^I_{(0)})}\Theta(X^I_{(0)})~\int _\Sigma
\partial^\alpha X^I 
\partial^\alpha y_i~
\label{bulksigma}
\end{equation} 
where the subscripts $(0)$ denote world-sheet zero modes.
Upon the interpretation of the Liouville zero mode $\phi_{(0)}$ as target
time, the deformations (\ref{bulksigma}) yield 
space-time metric deformations in 
a $\sigma$-model sense, which were interpreted in~\cite{ellis96}
as expressing the distortion of the space-time 
surrounding the recoiling $D$-brane soliton. 

For clarity,
we now drop the subscripts $(0)$ for the rest of this paper,
and we work in a region of space-time 
on the $D3$ brane such that $\epsilon (\phi - X^I)$ is finite 
in the limit $\epsilon \rightarrow 0^+$.   
The resulting space-time distortion is therefore
described by the metric elements 
\begin{equation} 
G_{0i} = (\epsilon^2 y_i + \epsilon u_i t)\Theta (t) + \epsilon g_{Ii}X^I \Theta (X^I)~,
\qquad i=1, \dots , m,~~I=m+1, \dots D-1
\label{gemteric}
\end{equation}  
to leading order in $\epsilon g_{Ii}$. 
The presence of $\Theta (t), \Theta(X^I)$ functions and
the fact that we are working in the region $y_i >0$
indicate that 
the induced space-time is piecewise continuous~\footnote{The 
important implications for non-thermal particle production
and decoherence for a spectator low-energy field theory
in such space-times were discussed in~\cite{kanti98,ellis96}, where only
the $D$--particle recoil case was considered.}.
In the general recoil/folding case considered in this article, 
the form of the resulting patch of the surrounding
space-time can be determined fully if one computes
the associated curvature tensors, along the lines 
of~\cite{kanti98}. 
Working to leading non-trivial order 
in the parameter $\epsilon$, we find, after some tedious 
but straightforward calculations, the 
following non-zero components of the Riemann tensor
(where $i\neq j$ and $I\neq J$):
\begin{eqnarray}
&~&R_{titi} =\epsilon^2 \delta (t) + \frac{1}{4} \sum _{I=m+1}^{D-1}
\epsilon^2 \left[ \Theta (X^I) + X^I\delta (X^I) \right] ^{2}
g_{Ii}^2 
+ {\cal O}(\epsilon^4) \nonumber \\
&~&R_{titj}=\frac{\epsilon^2}{4}\sum_{I=m+1}^{D-1} g_{Ii}g_{Ij}
\left[\Theta (X^I) + X^I \delta (X^I) \right]^2 \nonumber \\
&~&R_{titI} ={\cal O}(\epsilon ^3), \qquad R_{tiiI}={\cal O}(\epsilon ^4), 
\nonumber \\ 
&~&R_{tItI} =\frac{\epsilon^2}{4 } \sum _{j=1}^{m}g_{Ij}^2 
\left[ \Theta (X^I) + X^I \delta (X^I) \right]^2 + {\cal O}(\epsilon ^4) 
\nonumber \\
&~&R_{tItJ} =\frac{\epsilon^2}{4 } \sum _{j=1}^{m}g_{Ij}g_{Jj} 
\left[ \Theta (X^I) + X^I \delta (X^I) \right]
\left[ \Theta (X^J) + X^I \delta (X^J) \right]
+ {\cal O}(\epsilon ^4) 
\nonumber \\
&~&R_{tiIJ}={\cal O}(\epsilon^3)~, \qquad R_{ijij}={\cal O}(\epsilon ^4)~,
\nonumber \\
&~&R_{tIiJ}={\cal O}(\epsilon ^3)~,
\qquad R_{ijjI}={\cal O}(\epsilon ^3)~, \nonumber \\
&~&R_{tIiI}=-\frac{\epsilon}{2}g_{Ii}\left[\delta (X^I) 
+ X^I \delta '(X^I)\right] + {\cal O}(\epsilon ^3) \nonumber \\
&~&R_{ijIJ} =\frac{\epsilon^2}{4 } \left[g_{Ii}g_{Jj}-g_{Ij}g_{Ji}\right]
\left[\Theta (X^I) + X^I \delta (X^I) \right]\left[\Theta (X^J) 
+ X^J\delta (X^J)\right] + {\cal O}(\epsilon ^4) \nonumber \\
&~&R_{iIiI} =-\frac{\epsilon^2}{4 } g_{Ii}^2
\left[\Theta (X^I) + X^I \delta (X^I) \right]^2 + {\cal O}(\epsilon ^4)
\nonumber \\
&~&R_{iIjI} =-\frac{\epsilon^2}{4 } g_{Ii}g_{Ij}
\left[\Theta (X^I) + X^I \delta (X^I) \right]^2 + {\cal O}(\epsilon ^4)
\nonumber \\
&~&R_{iIiJ} =-\frac{\epsilon^2}{4 } g_{Ii}g_{Ji}
\left[\Theta (X^I) + X^I \delta (X^I) \right]
\left[\Theta (X^J) + X^J \delta (X^J) \right] + {\cal O}(\epsilon ^4)
\nonumber \\
&~&R_{iIjJ} =-\frac{\epsilon^2}{4 } g_{Ij}g_{Ji}
\left[\Theta (X^I) + X^I \delta (X^I) \right]
\left[\Theta (X^J) + X^J \delta (X^J) \right] + {\cal O}(\epsilon ^4)~.
\label{riemann}
\end{eqnarray} 
The scalar curvature is then:
\begin{eqnarray}
R & = & -2\epsilon ^2 m \delta (t) -
2\epsilon ^2 \sum _{i=1}^m \sum _{I=m+1}^{D-1} g_{Ii}[2\delta (X^I) 
+ X^I \delta '(X^I) ][(\epsilon y_i + u_i t)\Theta (t) 
+ g_{Ji} X^J \Theta (X^J)]  \nonumber \\
&~& -\frac{3\epsilon^2}{2} \sum _{i=1}^{m}
\sum _{I=m+1}^{D-1} g_{Ii}^2[\delta (X^I)
+ X^I\delta ' (X^I)]^2 + {\cal O}(\epsilon^3)~. 
\label {scalarcurv}
\end{eqnarray} 
Away from the defect hypersurface $\{ X^I=0 \}$,  
the resulting scalar curvature is negative and constant,
with the form:
\begin{equation} 
    R \simeq -\frac{3}{2}\epsilon ^2 \sum _{i=1}^{m}
\sum _{I=m+1}^{D-1} g_{Ii}^2 + {\cal O}(\epsilon ^4) 
\label{asympt}
\end{equation}
in the general $D$-brane/$D$-brane scattering case.
\vskip0.3cm

However, it is interesting to notice that 
the space is maximally symmetric,
in the sense of the curvature being given in terms of the 
metric tensor as: 
\begin{equation}
R_{abcd}={\cal K} \left(g_{ac}g_{bd} - g_{ad}g_{bc}\right)
\label{maxsym}
\end{equation}
{\it only } 
in the case that only one of the $g_{Ii}$ is non zero,
e.g., $g_{Xz}$. This describes
the situation depicted in Fig.~\ref{fig1}:
a $D1$ brane is moving along the $z$ direction,
perpendicular to a $D3$ brane
embedded in a $D$-dimensional space-time. 
In this case, we obtain the form (\ref{maxsym}) for the
Riemann tensor, with
\begin{equation}
{\cal K}=-\frac{\epsilon ^2}{4}g_{Xz}^2.
\label{maxsym2}
\end{equation}
The fact that the space is not in general maximally symmetric
has been confirmed by a detailed analysis of the general case,
whose details we do not discuss here.

In the special $D1$/$D3$ case, 
the resulting space-time away from 
the hypersurface $X=0$ looks, for $t,y_i >0$, 
like $AdS_3 \otimes {\rm E}_{D-3}$, where E$_{D-3}$ is a flat Euclidean
$(D-3)$-dimensional space.
String theory on AdS$_3$ has been the subject 
of many recent articles in the context of critical string or 
$D$--brane theory~\cite{ads3lit}. 
Much of the mathematical interest in string
theories formulated on such spaces lies in
the fact that such spaces have a correspondence
with well-known group structures based on $SL(2,R)$: in particular,
the boundary of the AdS$_3$ may be
formulated as an appropriate coset of $SL(2,R)$. This is
relevant to the study of many physical problems, such as
the quantum Hall effect and high-energy scattering in QCD~\cite{ellis99}, 
as
well as the two-dimensional 
black hole~\cite{WittenBH}. It is 
interesting and suggestive that a patch of the
space-time found here
resembles that structure. We also note that such a
target space-time corresponds to the
electrovac solution of gauged supergravity~\cite{electrovac},
which can be shown to leave space-time supersymmetry unbroken. 
\vskip0.3cm

We conclude with some
remarks about the metric (\ref{maxsym}), (\ref{maxsym2}).
\vskip0.3cm

\noindent 
(i) If the folding is ignored:
$g_{Xz}=0$, 
the 
space-time resembles flat Minkowski space-time
to ${\cal {O}}(\epsilon ^{2})$~\cite{kanti98}, 
upon making the following transformation for $t > 0$: 
\begin{equation} 
{\tilde z} = z + \frac{1}{2}\epsilon u_{z} t^2, \qquad  
{\tilde t}=t .
\label{rindler}
\end{equation} 
This implies that the space-time 
induced by the recoiling 
$D$ brane resembles,  for $t \gg 0$ and to order 
${\cal O}(\epsilon ^{2})$, 
a Rindler wedge space with `acceleration'
$\epsilon u_z$. This space is well known to 
produce a conical singularity when
the time is compactified to a Euclidean-signature `temperature', 
with deficit angle  
\begin{equation}
\delta_{0z} \sim 2\pi \left(1 - 1/\epsilon u_z \right)~.
\label{rindler2}
\end{equation} 
Such conical singularities break 
target space-time supersymmetry.
However, this particular conical singularity should be considered as
a `thermal' property of the space. 
Critical string theory in Rindler space
is poorly understood at present, and
the only case studied so far concerns the situation in which 
the deficit angle is quantized: 
$\delta_{0z}=2\pi(1-\frac{1}{N})$, where $N$ a positive integer,
in which case the situation is equivalent to a string moving on
a $Z_N$ orbifold~\cite{dabholkar}~\footnote{Even in 
this case, however, the perturbative superstring vacuum
is known to be plagued with tachyonic 
instabilities,
indicating that a non-perturbative analysis is in order.}.
This is obviously not the case in (\ref{rindler2}), given that
our recoil/folding analysis 
is valid for weakly-coupled string theory with $g_s \ll 1$,
and hence very massive D branes that move slowly:
$\epsilon |u_i| \ll 1$. The non-critical logarithmic conformal
field theory string approach we advocate
here may thus be of use in the generic study of strings in Rindler
spaces. 
\vskip0.3cm

\noindent(ii) A simpler (but also interesting) 
case in which a conical singularity is generated is that
in which the cone is 
produced by the $D3$ brane folding
alone, i.e., ignoring the recoil velocity $u_i \rightarrow 0^+$.
We look at the asymptotic structure of the 
distorted space-time obtained by such a folding of the $D$ brane 
along, say, the $Y$ axis, for $X \gg 0$ and $t \gg 0$,
suppressing the recoil contribution for simplicity.    
Upon performing the time transformation
$t \rightarrow t - \frac{1}{2}\epsilon g_{Xz} X z $, the 
line element of the above-mentioned asymptotic space-time 
becomes:
\begin{eqnarray} 
&~&ds^2 =-dt^2 + dy_{\perp}^2 + \left(1 
- \alpha^2~z^2\right)~dX^2 + \left(1 + \alpha ^2~X^2\right)~dz^2 
- 2\alpha~z~dX~dt~, \nonumber \\ 
&~&\alpha \equiv \frac{1}{2} \epsilon g_{Xz}~, 
\label{bendinglineel}
\end{eqnarray} 
where $y_{\perp}$ denotes collectively  
the remaining $D-3$ space-time coordinates.  
First of all, it is immediate to observe that 
there is no angular deficit for non-compact bulk dimension
$z$. This is in agreement with the maximally-symmetric 
AdS$_3$ structure,
and is in agreement with the above-mentioned fact 
that space-time supersymmetry is preserved by such space-times.

The situation changes drastically in
case where the 
extra dimension $z$ is {\it compact}, which is 
the case assumed in~\cite{dimopo98,antoniadis98,antoniadis99}.
For simplicity, we assume that $z$ lies on a circle $S^1$
of unit radius (in appropriate string units, in which all the 
other distances are measured). 
For compact $z$ and at fixed $X \sim 1/\epsilon \gg 0$ and $t \gg 0$,
such that $\alpha^2~X^2 \simeq g_{Xz}^2/4 $, 
we observe from the metric (\ref{bendinglineel})
that there exists a deficit angle 
in the circle around $z$:
\begin{equation} 
\delta \simeq (\pi g_{Xz}^2/4)
\label{deficitangle}
\end{equation}
implying the dynamical formation 
of a conical-like singularity 
due to the $D3$--brane folding in this asymptotic region of 
the bulk space-time.

Such singularities in general break bulk 
space-time supersymmetry~\cite{adrian+mavro99}.    
However, in view of the fact that the
folded $D$ brane is an excited state of 
the string/$D$--brane system, the phenomenon should be 
viewed as a symmetry obstruction rather
than a spontaneous breaking of symmetry, in the sense
that, although the ground state of the 
string/$D$--brane system is supersymmetric, recoil
produces a particular excited state
that does not respect that symmetry~\cite{witten95}.  
\vskip0.3cm

\noindent
(iii) As mentioned earlier, logarithmic field theory is the appropriate
tool for discussing transitions between different classical
string vacua described by different conformal field theories.
It has been shown previously, in the context of closed-string/$D$-particle
scattering, how one logarithmic pair of operators describes the
transition between flat space and an AdS space~\cite{kanti98,ellis98}, and
how another logarithmic pair describes transitions between different
two-dimensional string black holes~\cite{ellis96}. The logarithmic pair
discussed
here may open the way to a description of the transition between
flat space and a two-dimensional black hole, but
pursuing that possibility lies beyond the scope of this paper.

\section*{Acknowledgements}

The work of N.E.M. is partially supported by P.P.A.R.C. (U.K.). 
E.W. wishes to thank Oriel College (Oxford) for financial support and
the Department of Physics, University of Newcastle (Newcastle-upon-Tyne)
and CERN Theory Division for hospitality.

\end{document}